\documentclass[acmsmall]{acmart}

\usepackage{hyperref}
\usepackage{url}
\usepackage{comment}
\usepackage{multirow}
\usepackage{amsmath}
\usepackage{braket}
\newcommand{\Hsquare}{%
  \text{\fboxsep=-.2pt\fbox{\rule{0pt}{1ex}\rule{1ex}{0pt}}}%
}
\usepackage{mathrsfs}
\usepackage{xspace}
\usepackage{amsfonts}
\usepackage{mathtools}
\usepackage{listings}
\usepackage{xcolor}
\usepackage{fancyvrb}
\usepackage{framed}
\DeclareUnicodeCharacter{2212}{-}

\begin{document}

\title{QuDiet: A Classical Simulation Platform for Qubit-Qudit Hybrid Quantum Systems}

\author{Turbasu Chatterjee}
\email{turbasu.chatterjee@gmail.com}
\author{Arnav Das}
\email{arnav.das88@gmail.com}
\author{Subhayu Kumar Bala}
\email{balasubhayu99@gmail.com}
\affiliation{
  \institution{A. K. Choudhury School of Information Technology, University of Calcutta, Kolkata}
  \city{Kolkata}
  \country{India}
  \postcode{700106}
}
\author{Amit Saha}
\email{abamitsaha@gmail.com}
\affiliation{
  \institution{A. K. Choudhury School of Information Technology, University of Calcutta, Kolkata}
  \city{Kolkata}
  \country{India}
  \postcode{700106}
}
\affiliation{
  \institution{Atos, Pune, India}
  \country{India}
  \postcode{411045}
}

 \author{Anupam Chattopadhyay}
\email{anupam@ntu.edu.sg}
\affiliation{%
  \institution{School of Computer Science and Engineering, Nanyang Technological University, Singapore}
  \city{Singapore}
  \country{Singapore}
  }
  
\author{Amlan Chakrabarti}
\email{acakc@caluniv.ac.in}
\affiliation{%
  \institution{A. K. Choudhury School of Information Technology, University of Calcutta, Kolkata}
  \city{Kolkata}
  \country{India}
  \postcode{700106}
}

\begin{abstract}
  In the recent years, numerous research advancements have extended the limit of classical simulation of quantum algorithms. Although, most of the state-of-the-art classical simulators are only limited to binary quantum systems, which restrict the classical simulation of higher-dimensional quantum computing systems. Through recent developments in higher-dimensional quantum computing systems, it is realized that implementing qudits improves the overall performance of a quantum algorithm by increasing memory space and reducing the asymptotic complexity of a quantum circuit. Hence, in this article, we introduce \textbf{QuDiet}, a state-of-the-art user-friendly python-based higher-dimensional quantum computing simulator. \textbf{QuDiet} offers multi-valued logic operations by utilizing generalized quantum gates with an abstraction so that any naive user can simulate qudit systems with ease as compared to the existing ones. We simulate various benchmark quantum circuits in \textbf{QuDiet} and show the considerable speedup in simulation time as compared to the other simulators without loss in precision. Finally, \textbf{QuDiet} provides a full qubit-qudit hybrid quantum simulator package with quantum circuit templates of well-known quantum algorithms for fast prototyping and simulation. The complete code and packages of \textbf{QuDiet} is available at \href{https://github.com/LegacYFTw/QuDiet} {https://github.com/LegacYFTw/QuDiet} so that other platforms can incorporate it as a classical simulation option for qubit-qudit hybrid systems to their platforms.
  
\end{abstract}

\maketitle

\section{Introduction}

A significant progress has been made in quantum computing recently due to its asymptotic advantage over classical computing~\cite{nielsen_chuang_2010, Farhi_1998, Preskill_2018}. Moreover, with the introduction of the mathematical notion of a qudit in~\cite{Muthukrishnan_2000}, the boundaries of quantum computing are extended beyond the binary state space. Qudits are states in a $d$-dimensional Hilbert space where $d >2$, thus allowing a much larger state space to store and process information as well as simultaneous control operations~\cite{9410395, Wang_2020, qft, Bocharov_2017, Fan_2007, Khan_2006, Di_2013}. It has been shown that usage of qudits leads to circuit complexity reduction as well as enhanced efficiency of some quantum algorithms. For practical demonstrations, qudits have been realized on several different hardware, including photonic quantum systems~\cite{Gao_2020}, ion-trap systems \cite{qutrit}, topological quantum systems \cite{Cui_2015first, Cui_2015, bocharov2015improved}, superconducting systems \cite{PhysRevA.76.042319}, nuclear magnetic resonance systems \cite{Dogra_2014, Gedik_2015}, continuous spin systems \cite{Bartlett_2002, Adcock_2016} and molecular magnets \cite{Leuenberger_2001}. 

In order to continue innovation in the multi-valued logic space for quantum computing, an efficient, easy to use quantum computing simulator with the support of qudits is the need of the hour. Given the size of the unitaries in the qudit space, one also needs to consider the computational costs incurred while simulating such large complex systems, which presents a significant challenge even in the binary state-space simulators. The inception of this simulator was in the wake (or lack thereof) of accessible simulators that were capable of simulating multi-valued logic in an user-friendly manner. This meant that a lot of research time and effort was previously spent in the construction of matrices and checking their compatibility, dimensions and kronecker products manually before their output could be deciphered from a huge 1D-array \cite{cirq_developers_2022_6599601, sky, Qiskit}. As an example, let say a generalized gate is imposed on five qudits in a 4-dimensional quantum system. For simulation purpose, the matrix of that generalized gate of $4^5 \times 4^5$ $i.e.,$ $1024\times1024$ needs to be prepared manually, which apparently makes the simulation time-consuming and error-prone. 

Our proposed simulator, named \textbf{QuDiet}, claims to solve all that by providing suitable abstractions that shy the user away from behemoth calculations and focus on purely the logic building and quantum phenomenology in the higher dimensional space. \textbf{QuDiet} does this thanks to its simple, yet effective, lean architecture that could be used to debug implementations quickly and provide outputs without unnecessary computational or memory overhead. It also provides the users with the flexibility of adding gates accordingly to a quantum circuit with only a few commands.

\par The main contributions of this work are as follows: 
\begin{itemize}
    \item The first of its kind proposal for a simulator based on higher-dimensional state space, multi-valued logic, utilizing generalized quantum gates.
    \item Using sparse matrices and related algorithms at the core of all quantum operations to unlock potential speed-up.

    \item Using GPU acceleration and efficient memory maps to process large matrices with considerable speedup.
     \item Benchmarking multiple quantum circuits in qudit systems and showing overall simulation time for the different backends for the first time to the best of our knowledge.
    \item A full package with quantum circuit templates for fast prototyping and simulation.
\end{itemize}
The structure of this article is as follows. Section. \ref{2} describes the higher-dimensional quantum circuit and its classical simulation. Section. \ref{3} proposes the higher-dimensional quantum simulator, \textbf{QuDiet}. Section. \ref{4} analysis the efficiency of the proposed simulator with the help of benchmark circuits. Future scope of the proposed simulator is outlined in Section \ref{5}. Section. \ref{6} captures our conclusions.


\section{Preliminaries}\label{2}

In this section, firstly, we discuss about qudits and generalized quantum gates. Later we put some light on classical simulation of a higher-dimensional quantum circuit. 

\subsection{Higher-dimensional quantum circuits}

Any quantum algorithm can be expressed or visualized in the form of a quantum circuit. Commonly for binary quantum systems, logical qubits and quantum gates comprise these quantum circuits \cite{barenco}. The number of gates present in a circuit is called gate count and the number of qubits present in a circuit is known as qubit cost. In this work, we mainly deal with qudits and generalized quantum gates since our simulator is based on higher-dimensional quantum computing.

\subsubsection{Qudits}
A logical qudit that encodes a quantum algorithm's input/output in $d$-ary or multi-valued quantum systems is often termed as data qudit. Another sort of qudit used to store temporary findings is the ancilla qudit. The unit of quantum information in $d$-dimensional quantum systems is \textit{qudit}. In the~$d$ dimensional Hilbert space~$\mathscr{H}_d$, qudit states can be substantiated as a vector.

The vector space is defined by the span of orthonormal basis vectors $\{\ket0,\ket1,\ket2,\dots \ket{d-1}\}$. In qudit systems, the general form of quantum state can be stated as 
\begin{equation}
\ket{\psi}=\alpha_0 \ket0 +\alpha_1 \ket1 +\alpha_2 \ket2+\cdots+\alpha_{d-1} \ket{d-1}=
\begin{pmatrix}
\alpha_0 \\
\alpha_1 \\
\alpha_2 \\
\vdots   \\
\alpha_{d-1} \\
\end{pmatrix}
\end{equation}
where $|\alpha_0|^2+|\alpha_1|^2+|\alpha_2|^2+\cdots+|\alpha_{d-1}|^2=1$ and $\alpha_0$, $\alpha_1$, $\dots$, $\alpha_{d-1} \in\mathbb{C}^d$.

\subsubsection{Generalized Quantum Gates}

In this section, a brief discussion on generalized qudit gates is exhibited. The generalization can be described as discrete quantum states of any arity in this way. Unitary qudit gates are applied to the qudits to evolve the quamtum states in a quantum algorithm. It is required to take into account one-qudit generalized gates such as NOT gate ($X_d$), Phase-shift gate ($Z_d$), Hadamard  gate ($F_d$), two-qudit generalized CNOT gate ($C_{X,d}$) and generalized multi-controlled Toffoli gate ($C^{n}_{X,d}$) for logic synthesis of quantum algorithms in $d$-dimensional quantum systems. For better understanding, these gates are described in detail:

\textbf{Generalized NOT Gate:} $X^d_{+a}$, the generalized NOT can be defined as $X^d_{+a}\ket{x}=\ket{(x+a) \mod d}$, where $1 \le a \le d-1$. For visualization of the $X^d_{+a}$ gate, we have used a 'rectangle' ($\Hsquare$). '$X^d_{+a}$' in the 'rectangle' box represents the generalized NOT.

\textbf{Generalized Phase-Shift Gate}

$Z_d$ is the generalized phase-shift gate represented by a $(d \times d)$ matrix is as follows, with $\omega=e^{\frac{2\pi i}{d}}$ henceforth:

  \begin{align*}
Z_d = \left(\begin{matrix} 1 & 0 & 0 & \ldots & 0 \\ 0 & \omega & 0 & \ldots & 0 \\ 0 & 0 & \omega^2 & \ldots & 0 \\ \vdots & \vdots & \vdots & \ddots & \vdots \\ 0 & 0 & 0 & \ldots & \omega^{d-1}\end{matrix}\right)
  \end{align*}
  
We have used $Z_d$ in the 'rectangle' ($\Hsquare$) box to represent the generalized phase-shift gate.

\textbf{Generalized Hadamard Gate:} The superposition of the input basis states is produced via the generalized quantum Fourier transform, also known as the generalized Hadamard gate, $F_d$. The generalized quantum Fourier transform or generalized Hadamard gate, produces the superposition of the input basis states. We have used $F_d$ in the 'rectangle' ($\Hsquare$) box to represent the generalized Hadamard gate. The $(d \times d)$ matrix representation of it is as shown below  
:

\begin{align*}
F_d = {1\over\sqrt{d}} \left(\begin{matrix} 1 & 1 & 1 & \ldots & 1 \\ 1 & \omega & \omega^2 & \ldots & \omega^{d-1}  \\ 1 & \omega^2 & \omega^4 & \ldots & \omega^{2(d-1)}  \\ \vdots & \vdots & \vdots & \ddots & \vdots \\ 1 & \omega^{d-1} & \omega^{2(d-1)} & \ldots & \omega^{(d-1)(d-1)} \end{matrix}\right)
  \end{align*}

\textbf{Generalized CNOT Gate:} In a binary quantum system, a controlled NOT (CNOT) gate can achieve quantum entanglement, which is an unrivalled property of quantum mechanics. For $d$-dimensional quantum systems, the binary two-qubit CNOT gate is generalised to the $INCREMENT$ gate:\\ $\text{INCREMENT}\ket{x}\ket{y}=\ket{x}\ket{(x+a) \mod d}$, if $x=d-1$, and = $\ket{x}\ket{y}$, otherwise, where $1 \le a \le d-1$. In schematic design of the generalized CNOT gate, $C_{X,d}$, we have used a 'Black dot' ($\bullet$) to represent the control, and a 'rectangle' ($\Hsquare$) to represent the target. '$X^d_{+a}$' in the target box represents the increment operator. \\ The $(d^2 \times d^2)$  matrix representation of the generalized CNOT $C_{X,d}$ gate is as follows:

\begin{equation*}
C_{X,d} = \left( \begin{matrix}
    I_d & 0_d & 0_d & \ldots & 0_d \\
    0_d & I_d & 0_d & \ldots & 0_d \\
    0_d & 0_d & I_d & \ldots & 0_d \\
    \vdots & \vdots & \vdots & \ddots & \vdots \\
    0_d & 0_d & 0_d & \ldots &  X^d_{+a} \\
\end{matrix} \right)
  \end{equation*}

where $I_d$ and $0_d$ are both $d \times d$ matrices as shown below:
\begin{equation*}
I_d = 
\begin{pmatrix}
    1 & 0 & 0 & \ldots & 0 \\
    0 & 1 & 0 & \ldots & 0 \\
    0 & 0 & 1 & \ldots & 0 \\
    \vdots & \vdots & \vdots & \ddots & \vdots \\
    0 & 0 & 0 & \ldots &  1 \\
\end{pmatrix} 
\quad\textrm{and,}\quad
0_d =  
\begin{pmatrix}
    0 & 0 & 0 & \ldots & 0 \\
    0 & 0 & 0 & \ldots & 0 \\
    0 & 0 & 0 & \ldots & 0 \\
    \vdots & \vdots & \vdots & \ddots & \vdots \\
    0 & 0 & 0 & \ldots &  0 \\
\end{pmatrix}
\end{equation*}

\textbf{Generalized Multi-controlled Toffoli Gate:} We expand the generalized CNOT or $INCREMENT$ further to work over $n$ qudits as a generalized Multi-controlled Toffoli Gate or $n$-qudit Toffoli gate $C_{X,d}^n$.  For $C_{X,d}^n$, the target qudit is increased by $a \ (\text{mod } d)$ only when all $n-1$ control qudits have the value $d-1$, , where $1 \le a \le d-1$. In schematic design of the generalized Multi-controlled Toffoli Gate, $C_{X,d}^n$, we have used 'Black dots' ($\bullet$) to represent all the control qudits, and a 'rectangle' ($\Hsquare$) to represent the target. '$X^d_{+a}$' in the target box represents the increment operator. The $(d^n \times d^n)$ matrix representation of generalized Multi-controlled Toffoli (MCT) gate is as follows:

\begin{equation*}
C_{X,d}^n = \left( \begin{matrix}
    I_d & 0_d & 0_d & \ldots & 0_d \\
    0_d & I_d & 0_d & \ldots & 0_d \\
    0_d & 0_d & I_d & \ldots & 0_d \\
    \vdots & \vdots & \vdots & \ddots & \vdots \\
    0_d & 0_d & 0_d & \ldots &  X^d_{+a} \\
\end{matrix} \right)
  \end{equation*}

\begin{figure}[!htb]
\centering
\includegraphics[scale=.7]{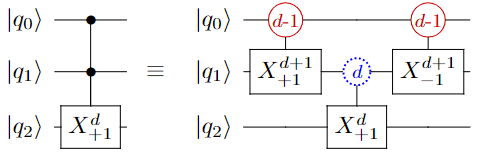}
\caption{Generalized Toffoli in $d$-ary quantum systems --- the control qudits in red circles  activate on $\ket{d-1}$
and those in the blue  dotted-circles  activate on $\ket{d}$.}
\label{gentofdec}
\end{figure}

For the sake of simplicity, we decompose generalized Multi-controlled Toffoli gate into set of generalized CNOT gates in rest of the article \cite{Gokhale_2019, sahapra}. It is also exhibited that this decomposition of Toffoli gate can be logarithmic in depth as compared to linear depth while using conventional approach of using generalized $T$ gate. This depth reduction is also useful for implementing different algorithms in quantum computing. A generalized Toffoli decomposition in a $d$-ary system using $\ket{d}$  state is shown in Figure \ref{gentofdec}.

\subsection{Classical simulation of a higher-dimensional quantum circuit}

This section highlights how a program that runs on a classical computer can resemble the evolution of a quantum computer. Before jumping on that let us discuss about the challenges that the current qubit-only classical simulators are facing to simulate qudit systems or higher-dimensional quantum circuits.
\begin{itemize}
    \item For the state-of-the-art qubit-only simulators \cite{cirq_developers_2022_6599601, Qiskit, Steiger2018projectqopensource, rigetti, liquid}, the simulators need to act on an $n$ qubit state with a $2^n\times2^n$ matrix. The dimension of the matrices of the unitary gates are also quite straight-forward as it is only based on qubit systems \cite{LaRose_2019}. Due to the engineering challenge of maintaining the dimension of the matrices automatically for the qubit-qudit hybrid systems, the current simulators are unable to provide a solution to simulate the higher-dimensional quantum circuits efficiently, which is addressed in this paper.
    
    \item The other challenge is that it requires a significant amount of memory to store the higher-dimensional quantum state vectors and to perform matrix multiplication to simulate the generalized quantum gates. Hence, the current simulators reach long simulation times and memory limitations very quickly due to its conventional memory management. In this paper, we also address this issue by designing unitary matrix simulator with various backends to simulate qudit systems effectively.
\end{itemize}
  Before explaining our proposed simulator more elaborately, we would like to discuss about the technicalities of the classical simulation of a higher-dimensional quantum circuit.   

To simulate a higher-dimensional quantum circuit, we first need to specify the dimension of each qudit so that generalized quantum gates or quantum operations can act on a sequence of qudits effectively~\cite{cirq_developers_2022_6599601}. This can be done through a method, which returns a tuple of integers corresponding to the required dimension of each qudit it operates on, as an instance (2, 3, 4) means an object that acts on a qubit, a qutrit, and a ququad. To apply a generalized gate to some qudits, the dimensions of the qudits must match the dimensions it works on. For example, for a single qubit gate, its unitary is a $2\times2$ matrix, whereas for a single qutrit gate its unitary is a $3\times3$ matrix. A two qutrit gate will have a unitary that is a $9\times9$ matrix $(3 \times 3 = 9)$ and a qubit-ququad gate will have a unitary that is an $8\times8$ matrix $(2 \times 4 = 8)$. The size of the matrices involved in defining mixtures and channels follow the same pattern.

After simulating higher-dimensional quantum circuit by considering the dimension of qudits and generalized gates appropriately, the size of the resultant state is determined by the product of the dimensions of the qudits being simulated. For example, the state vector output after simulating a circuit on a qubit, a qutrit, and a ququad will have $2 \times 3 \times 4 = 24$ elements, Since, circuits on qudits are always assumed to start in the  computational basis state $\ket{0}$, and all the computational basis states of a qudit are assumed to be $\ket{0}, \ket{1}, \dots, \ket{d-1}$. Measurements of qudits are assumed to be in the computational basis and for each qudit return an integer corresponding to these basis states. Thus measurement results for each qudit are assumed to run from $\ket{0}$ to $\ket{d}$ like $\ket{0}$ to $\ket{1}$ for qubit systems.

\section{\textbf{QuDiet}: A Qubit-Qudit Hybrid Quantum Simulator}\label{3}
\begin{figure*}[h!]
    \centering
    \includegraphics[scale=0.36]{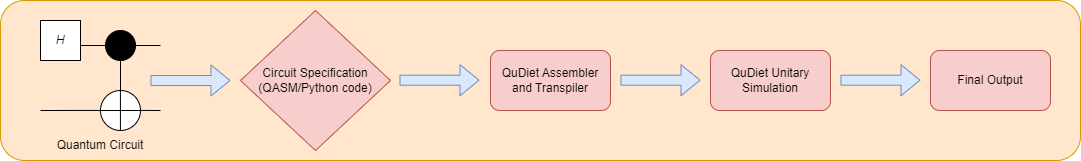}
    \caption{A user-level compilation flow of the \textbf{QuDiet}, where the input is a quantum circuit and the output is the probable quantum states of the input circuit.}
    \label{fig:user}
\end{figure*}

The lean architecture of \textbf{QuDiet} has been laid out briefly in the subsequent subsections. Before that the flow on the user end can be summed in Figure \ref{fig:user}. This figure shows the high-level description of \textbf{QuDiet} to understand the general features of the proposed simulator. First, the quantum algorithm is expressed as a quantum circuit with the help of either QuDiet's QASM specification or simple python console. Next, this needs to be compiled to a specific quantum gate set of \textbf{QuDiet}. Finally, the quantum circuit is simulated with \textbf{QuDiet} to get the final outcome of the given quantum algorithm.

Now that we have a high level understanding of the compiler, let's take a deep dive and look at the internals in the following subsections.

\subsection{High level architecture}
\begin{figure}[h!]
    \centering
    \includegraphics[scale=0.33]{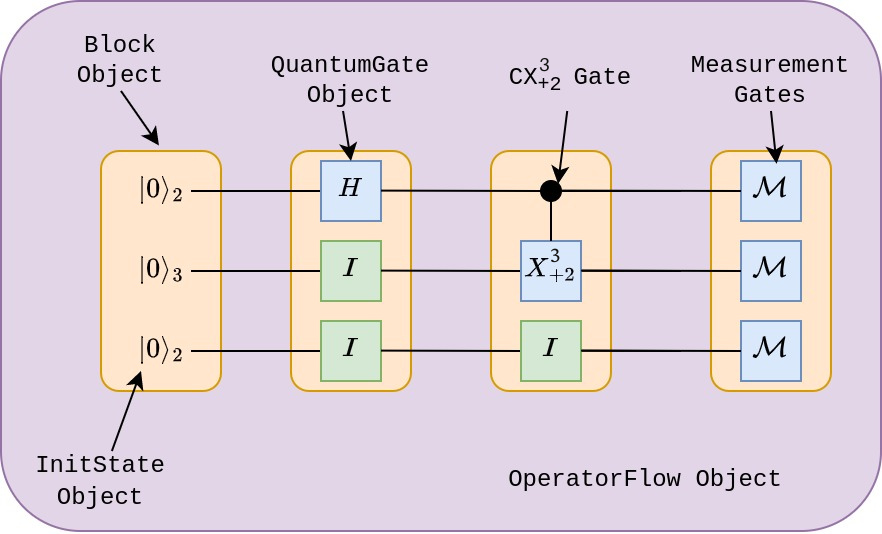}
    \caption{High level overview of the \textbf{QuDiet} quantum simulator. This figure shows the different parts of the quantum simulator, namely the \texttt{InitState} objects, with the subscripts representing the dimensions of the respective qudits, and the \texttt{QuantumGate} objects with the $CX_{+2}^3$ gate being a generalized version of the CNOT gate with the target acting on a qutrit whereas the control being a qubit. The \texttt{QuantumGate} objects are enclosed using the \texttt{Moment} object and the \texttt{Moment} objects are enclosed within the \texttt{OperatorFlow} object.}
    \label{fig:my_label}
\end{figure}

At its core, \textbf{QuDiet} is managed by two integral parts: The \texttt{Moment} and the \texttt{OperatorFlow} objects. These two, however, are simple vector objects that behave like a stack, during the compiler's operation. This is portrayed in Figure \ref{fig:my_label}. A circuit, in essence, is an \texttt{OperatorFlow} object at its heart, running on a \texttt{Backend}. Once a \texttt{QuantumCircuit} object has been instantiated, in the background an \texttt{OperatorFlow} object is also created, which consists of a single \texttt{Moment} object. This \texttt{Moment} object can carry two types of objects: A \texttt{InitState}, which is the representation of an initial state of a quantum circuit or an \texttt{QuantumGate}, an abstract class, inherited by all quantum gates that are implementable by the simulator.

Once the user has implemented a quantum circuit, using the available commands, the \texttt{measure\_all()} function is invoked, thereby pushing a \texttt{Moment} carrying measurement gates to all quantum registers. The measurement gate is a symbolic gate that tells the compiler that a program routine has ended and can be executed. The execution occurs when the \texttt{QuantumCircuit.run()} method is invoked, using the circuit's specified \texttt{Backend} object which has several interfaces for plug and play operability. 

\subsection{The Quantum Circuit}
The quantum circuit is represented by the \texttt{QuantumCircuit} class in the simulator. Whenever a new quantum circuit is invoked, a QuantumCircuit object is instantiated. This \texttt{QuantumCircuit} object takes the following arguments for instantiation: 

\begin{itemize}
    \item \texttt{qregs}: The dimensions of the quantum registers is represented as a heterogeneous list of integer dimensions. In other words, the dimension of the quantum `wire', in order, or as a tuple represents a homogeneous register of fixed length, with the first qudit acting as the Least Significant Bit (LSB) and the last qudit in the register acting as the Most Significant Bit (MSB). 
    \item \texttt{cregs}: The length of the classical register (optional)
    \item \texttt{name}: A string that represents the name of the quantum circuit (optional)
    \item \texttt{init\_states}: Represents the configuration of the initial states of the register. This is represented by an array of the same length as that of \texttt{qregs}. If none is provided, the registers gets automatically initialized to $|0\rangle$'s.
    \item \texttt{backend}: This represents the \texttt{Backend} on which the quantum circuit is to be executed. There are four \texttt{Backend} objects to choose from, and are elaborated in subsection \ref{backends}. The default backend to be used is the \texttt{SparseBackend}
    \item \texttt{debug}: A flag argument that forms the base of a debugger engine, implemented in a simple manner in this release cycle and shall be expanded upon in future versions for easy debugging and callback functions.
\end{itemize}

For example, in order to make a circuit with 3 qudits of dimensions 4,5,3 respectively, with initial states being $|0\rangle$, $|3\rangle$ and $|2\rangle$, that calculates using the \texttt{CUDASparseBackend} we just need the following lines of \texttt{python3} code: \\
\texttt{qreg\_dims = [4,5,3]}\\
\texttt{init\_states = [0,3,2]}\\
\texttt{backend = CUDASparseBackend}\\
\texttt{qc = QuantumCircuit(qregs=qreg\_dims, init\_states=init\_states, backend=backend)}

This shows how easily we can simulate different dimensional qudits with this simulator. Let's now take a closer look at these initial states, or more generally, the quantum states that they represent and how does \textbf{QuDiet} handle them.

\subsection{Representation of quantum states}
Any quantum simulator is incomplete without their interpretation of quantum states. This preliminary version of \textbf{QuDiet} assumes that quantum states as state vectors. This comes with a caveat that \textbf{QuDiet} can only deal with states, as represented by an array or a vector list. This will of course be improved upon in future releases where we hope to incorporate density matrices, tensor-network and ZX/ZH calculus based representations. But we shall stick to the notion that these state vectors would be represented by: 

\begin{equation}
\ket{\psi}=\alpha_0 \ket0 +\alpha_1 \ket1 +\alpha_2 \ket2+\cdots+\alpha_{d-1} \ket{d-1}=
\begin{pmatrix}
\alpha_0 \\
\alpha_1 \\
\alpha_2 \\
\vdots   \\
\alpha_{d-1} \\
\end{pmatrix}
\end{equation}

Therefore, a register of qudits $|\psi_1 \psi_2 \dots \psi_n \rangle$ would now need to be represented as:

\begin{equation}
\begin{split}
    |\psi_1\rangle \otimes \dots \otimes |\psi_n\rangle &= 
  \begin{pmatrix}
\alpha_{10} \\
\alpha_{11} \\
\alpha_{12} \\
\vdots   \\
\alpha_{1 d-1} \\
\end{pmatrix}
\otimes
\dots
\otimes
\begin{pmatrix}
\alpha_{n0} \\
\alpha_{n1} \\
\alpha_{n2} \\
\vdots   \\
\alpha_{n d-1} \\
\end{pmatrix}  
\end{split}
\end{equation}

Unlike qubit-only quantum simulators, this presents an engineering challenge: An efficient method of storing these matrices and their computations in memory. On observation, we note that the number these quantum states are, in essence very sparse matrices with non-zero elements scattered around them. Therefore, it seemed natural to use the sparse array format for storing these state vectors. 

In order to achieve this, \textbf{QuDiet} makes use of \texttt{scipy}'s Compressed Sparse Column array implementation. We shall be improving on this format to reduce latency of Sparse Matrix Vector multiplication (SpMV) and General Matrix Vector multiplication (GeMV). This is elaborated in the Section V. However we have added support for \texttt{numpy} matrices for small circuits.

This technique is also used when representing quantum operators or quantum gates as we shall see in the following subsection.

\subsection{Representation of quantum gates}

One of the key features of \textbf{QuDiet} is it's ability to construct generalized gates and operators for quantum computing using multi-valued logic automatically. This is a standout feature and there does not exist a simulator in known literature that allows the construction of single and multi-qudit quantum gates and operators with the ease as that of \textbf{QuDiet}. 

The \textbf{QuDiet} contains a limited gate set. These are as follows, and their descriptions are available in the Section \ref{2}: 
\begin{enumerate}
    \item NOT Gate (XGate)
    \item Phase-Shift Gate (ZGate)
    \item Hadamard Gate (HGate)
    \item CNOT Gate (CXGate)
    \item User Defined Gate (QuantumGate)
    \item Measurement Gate 
    \item Identity Gate (IGate)
\end{enumerate}

Each of these quantum gates take into account two things: The qudit register it is acting on, and the dimension of the qudit register. Using a user-defined acting register, these quantum gates are able to take into account the dimension of the acting register, and dynamically construct a gate unitary at runtime. The XGate and the CXGate have an added functionality, that an arbitrary shift can be induced, as long as the shift has a value less than the dimension of the qudit register. The Measurement Gate in \textbf{QuDiet} is a special gate in that, it has no unitary associated with it. Up until this version, the Measurement Gate merely acts as flag that signifies the end of a quantum circuit. Any gate post measurement will be ignored by the simulator.

Once the unitary of a generic quantum gate has been created, it then utilizes the same sparse matrix format to store the data. Contrary to quantum states, however, quantum gates utilize \texttt{scipy}'s Compressed Sparsed Row matrix implementation. This format will of course, be improved upon to not only facilitate SpMV and GeMV but also SpGeMM or Sparse Matrix-Matrix multiplication and GeMM, or General Matrix-Matrix multiplication \cite{sparse}. However we have added support for \texttt{numpy} matrices for small circuits.

The backends have been engineered to provide minimal speedup using naive algorithms, and has CUDA support for GPU executions. These backends are elaborated in subsection \ref{backends}.

Therefore for demonstration, if we were to use the same quantum circuit as previously, in order to add a Hadamard gate acting on the first qudit and a CNOT gate with a shift of plus 2, acting on the first and the third qudit, we simply invoke the following lines of python3 code: \\
\texttt{qc.h(0)} \\
\texttt{qc.cx((0,2), plus=2)}

\textbf{QuDiet} also decomposes Toffoli gates (Let's say, \texttt{qc.Toffoli((0,1,2), plus=1}) into suitably mapped higher order \texttt{CXGate} objects. This is given as follows: \\ 
\texttt{qc.cx((0,1), plus=1)}\\
\texttt{qc.cx((1,2), plus=1)}\\
\texttt{qc.cx((0,1), plus=2)}

As stated before, these form the basis of a \texttt{Moment} object, which we shall elaborate on next.

\subsection{The Moment}
The \texttt{Moment} object is a forms an abstraction between the \texttt{QuantumGate} and the \texttt{OperatorFlow} objects, in that it maintains the orientation of the quantum gates that are acting on the respective qudits and the \texttt{OperatorFlow} maintains the sequence of execution of the quantum operations. The \texttt{Moment} object is inherently an array of length equal to the breadth of the quantum circuit. The primary job of the \texttt{Moment} object is to maintain the position of an acting register on the intended qudit so that errors in evaluating Kronecker products are avoided when the quantum circuit is executed.

There can be a single \texttt{Moment} object containing a list of  \texttt{InitState} objects spanning across all the qudits in the quantum register and is initialized and pushed into the \texttt{OperatorFlow} object's stack whenever a quantum circuit is initialized. The \texttt{OperatorFlow} object can hold an arbitrary number of \texttt{Moment} objects, containing quantum gates across the breadth of the quantum register. 

As per the \textbf{QuDiet}, whenever the user invokes a quantum gate onto a quantum circuit, the \texttt{QuantumGate} is pushed into the specific register corresponding to the index in the \texttt{Moment} object. All other corresponding registers or indices in the \texttt{Moment} object shall contain the \texttt{IdentityGate} object. 

Something to note here is that, the \texttt{OperatorFlow} and the \texttt{Moment} data structures only store the data when it is pushed. No kronecker product or matrix multiplication operation would be done until the Measurement gates would be pushed and the user invokes the \texttt{run()} method from the quantum circuit object. 

\textbf{QuDiet} also performs preliminary optimizations at the logic level whenever a gate is pushed. Whenever a new gate is pushed into the \texttt{OperatorFlow} stack, the gate is enclosed in a \texttt{Moment} object, while ensuring that an index in the \texttt{Moment} array corresponds to the acting qudit as specified by the user. All the other registers have an Identity gate acting on them. When pushing a \texttt{Moment} object containing a quantum gate into the \texttt{OperatorFlow} stack, the simulator checks if the any other immediate predecessor \texttt{Moment} has an Identity gate in them at the same position, if so, the Identity Gate is swapped out for the currently incoming quantum gate. This is done until it reaches an \texttt{InitState} object or it finds a \texttt{QuantumGate} object in any of its immediate predecessors.

\subsection{The Operator Flow Stack}
The \texttt{OperatorFlow} object is at the heart of the \textbf{QuDiet}: it maintains the order of execution of \texttt{QuantumGate} objects nestled inside the \texttt{Moment} objects. The \texttt{OperatorFlow} inherently maintains a vector list which acts like a stack during execution. 

In order to run a quantum circuit, a \texttt{measure\_all()} is called. This places \texttt{MeasurementGate} objects, across all quantum registers, thereby raising flag variables inside a \texttt{Moment} object. Any \texttt{Moment} containing quantum gates pushed after the \texttt{Moment} object shall be discarded, post the \texttt{measure\_all()} command. In order to execute the said circuit, the \texttt{run()} is invoked, thereby outputting the statevectors of the quantum states that have non-negative probabilities. 

Under the hood, during this time, all circuits prior to the \texttt{Moment} containing the \texttt{MeasurementGate} objects are called to be executed in reverse order. This means that the \texttt{OperatorFlow} object will first `pop' out that \texttt{Moment} and evaluate the Kronecker product depending on the type of \texttt{Backend}, inside the \texttt{Moment} and store it in a variable. Then it will advance onto the second-to-last \texttt{Moment}, evaluate the Kronecker product, and then store it in a separate variable. Now once these two Kronecker products have been evaluated, \textbf{QuDiet} will perform a SpGEMM or a GEMM operation, depending on the \texttt{Backend} selected for the circuit. When the matrix products have been evaluated, the two variable storing the Kronecker product is freed from memory. This continues until the \texttt{Moment} containing the \texttt{InitState} objects are reached where the last operation is and SpMV or a GEMV, depending on the \texttt{Backend} chosen.

\subsection{Acceleration and the Backends} \label{backends}

In \textbf{QuDiet}, acceleration can be achieved in two different ways. One is through a GPU, i.e., hardware acceleration, and the other is by using sparse matrices, i.e., algorithmic or software acceleration. These accelerations are delivered through different Backends, like \textbf{CudaBackend}. To access the GPU as a host for hardware acceleration, we use \textbf{CuPy}, which is a GPU equivalent for \textbf{NumPy} and \textbf{SciPy}. In terms of software acceleration, \textbf{SciPy} is used instead of \textbf{NumPy}, because of its useful interface for Sparse matrices. The very basis of \textbf{QuDiet}, lies in the availability and choice of Backends. For example, nearly dense matrices have showed no acceleration when run on the \textbf{SparseBackend}. On the contrary, it is a better choice to use \textbf{CudaBackend} instead and ignore the \textbf{sparse like Backends} when dealing with nearly dense matrices. A more detail discussion on Backends is carried out in next. 

\textbf{QuDiet} is a model-level library, providing high-level building blocks for developing quantum circuit algorithms. The low-level operations such as dot product, kronecker product, etc. These low-level operations are interfaced through the class \texttt{Backend}. This enables one to use the best backend option based on the type of the circuit and the hardware accessible.

Right now, the backends accessible for use are \textbf{NumpyBackend}, \textbf{SparseBackend}, \textbf{CudaBackend} and \textbf{CudaSparseBackend}

\subsubsection{NumpyBackend}
NumpyBackend is the default backend used when no backend type is explicitly defined. This interfaces to the basic numpy operations, without any external optimization.

\subsubsection{SparseBackend}
SparseBackend interfaces to scipy's sparse module instead of interfacing to numpy's ndarray. It stores only the nonzero elements of the matrix and
reduce the computation time by eliminating operations on zero elements. In cases, this can reduce the matrix size exponentially, showing an overall speedup in execution runtime and memory compression.

This speedup directly depends on the nature of the circuit, where the worst scenario is the arrays being mostly dense. In that case, the SparseBackend will perform nearly like a NumpyBackend.

\subsubsection{CUDABackend}
CudaBackend interfaces with \textbf{CuPy's Numpy Routine}, using numpy-like dense matrices, accessing the GPUs with the purpose of runtime speedup only.

\subsubsection{CUDASparseBackend}
CUDASparseBackend interfaces with \textbf{CuPy's Scipy Routine}, using sparse representation of the matrices and then accessing the GPUs with the purpose of runtime speedup.

The CUDASparseBackend optimizes the operations using sparse matrices, followed by hardware(\textbf{GPU}) optimization.

\subsection{Output and Interpretability of results}

Output Representation of a quantum circuit is still a less explored domain, especially when dealing with large circuits. \textbf{QuDiet} additionally provides some small contributions in terms of Output Representation and Interpretability for the sake of better research experience. \textbf{QuDiet} comes with two types of output representation, \textsc{OutputType} and two types of output method, \textsc{OutputMethod}. \textsc{OutputType} is the way of output representation, which has two types, \textbf{print} and \textbf{state}. The \textsc{OutputType.state} provides the raw output state as a binary array. Whereas the \textsc{OutputType.print} provides the output state as a \textit{ket} string. On the contrary, \textsc{OutputMethod} dictates whether the output would provide the probability distribution or the amplitude of the quantum states. By default, a quantum circuit returns the final \textit{ket} representation along with the distribution probability, which looks like \textsc{\{$\ket{1110100000000} : 1.0$ \}} for an instance.


\subsection{Example Workflow}
\begin{figure}[!ht]
\centering
\includegraphics[scale=.2]{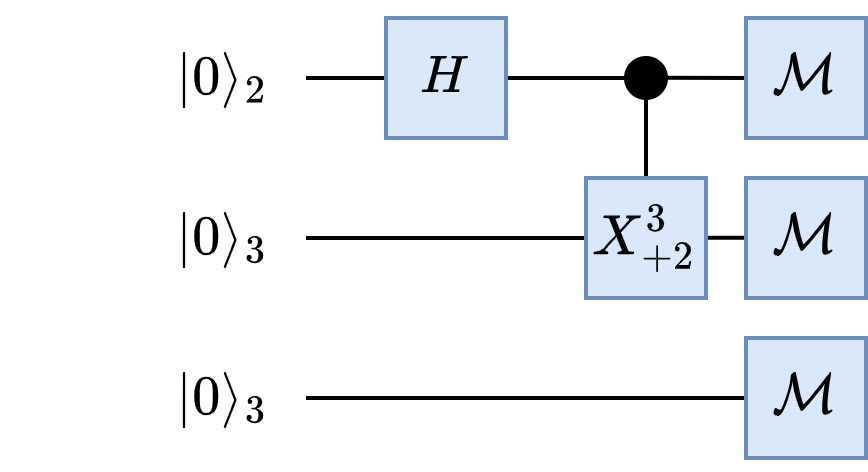}
\caption{A quantum circuit to be simulated using \textbf{QuDiet}}
\label{examplecirc}
\end{figure}

Let us now, take an example to understand the flow of work in \textbf{QuDiet}. The circuit to be executed is shown in Figure \ref{examplecirc}. 

In order to do this, we must first create the quantum circuit by specifying the dimensions of the circuit lines in the quantum circuit, this is done by the following lines of python code: \\
\texttt{qreg\_dims = [2,3,3]}\\
\texttt{init\_states = [0,0,0]}\\
\texttt{backend = SparseBackend}\\
\texttt{qc = QuantumCircuit(qregs=qreg\_dims, init\_states=init\_states, backend=backend)}

These lines of code do the following: 
\begin{enumerate}
    \item The \texttt{InitState} objects are created for each of the circuit lines. Each of these \texttt{InitState} objects have the following matrices: $$ \begin{pmatrix}
1 \\
0 \\
 
\end{pmatrix},   \begin{pmatrix}
1 \\
0 \\
0
 
\end{pmatrix} \textrm{, and}
\begin{pmatrix}
1 \\
0 \\
0
 
\end{pmatrix}$$ for the initiated circuit lines respectively.
    
    \item The \texttt{Moment} object is initialized and then the \texttt{InitState} objects are pushed into the \texttt{Moment}. This object is then pushed into the \texttt{OperatorFlow} object's stack.
\end{enumerate}

Now that the circuit has been initialized with the desired states, we can now add in the required gates. These gates are invoked by calling the respective methods of the circuit, which then creates the respective objects of the quantum gates. These quantum gates, at the time of their creation, take into account the dimensions of the acting register, among other factors, to construct the correct unitary automatically and push it into the \texttt{Moment} object, which is then pushed into the \texttt{OperatorFlow} object. This is done by the following lines of code: \\
\texttt{qc.h(0)}\\
\texttt{qc.cx((0,1), plus=2)}\\

These lines of code does the following: 
\begin{enumerate}
    \item The first line of the above snippet detects the dimension of the acting register. Since the dimension of the acting register is 2, it shall construct a $2 \times 2$ unitary suitably as follows: $$\begin{pmatrix}
        1/\sqrt{2} && 1/\sqrt{2} \\
        1/\sqrt{2} && -1/\sqrt{2}
    \end{pmatrix}$$
    Once done, it will assign it to the instance variable of the \texttt{HGate} object. 
    \item Before pushing the \texttt{HGate} object to the \texttt{Moment} list, \textbf{QuDiet} will create Identity gates tailored to the dimensions of the qudit register and push it into the \texttt{Moment} object.
    \item Next, \textbf{QuDiet} will create the \texttt{CXGate} object using a generated unitary and then place the Identity gates before pushing it into the \texttt{Moment} object.
\end{enumerate}

\begin{figure}[!ht]
\centering
\includegraphics[scale=.091]{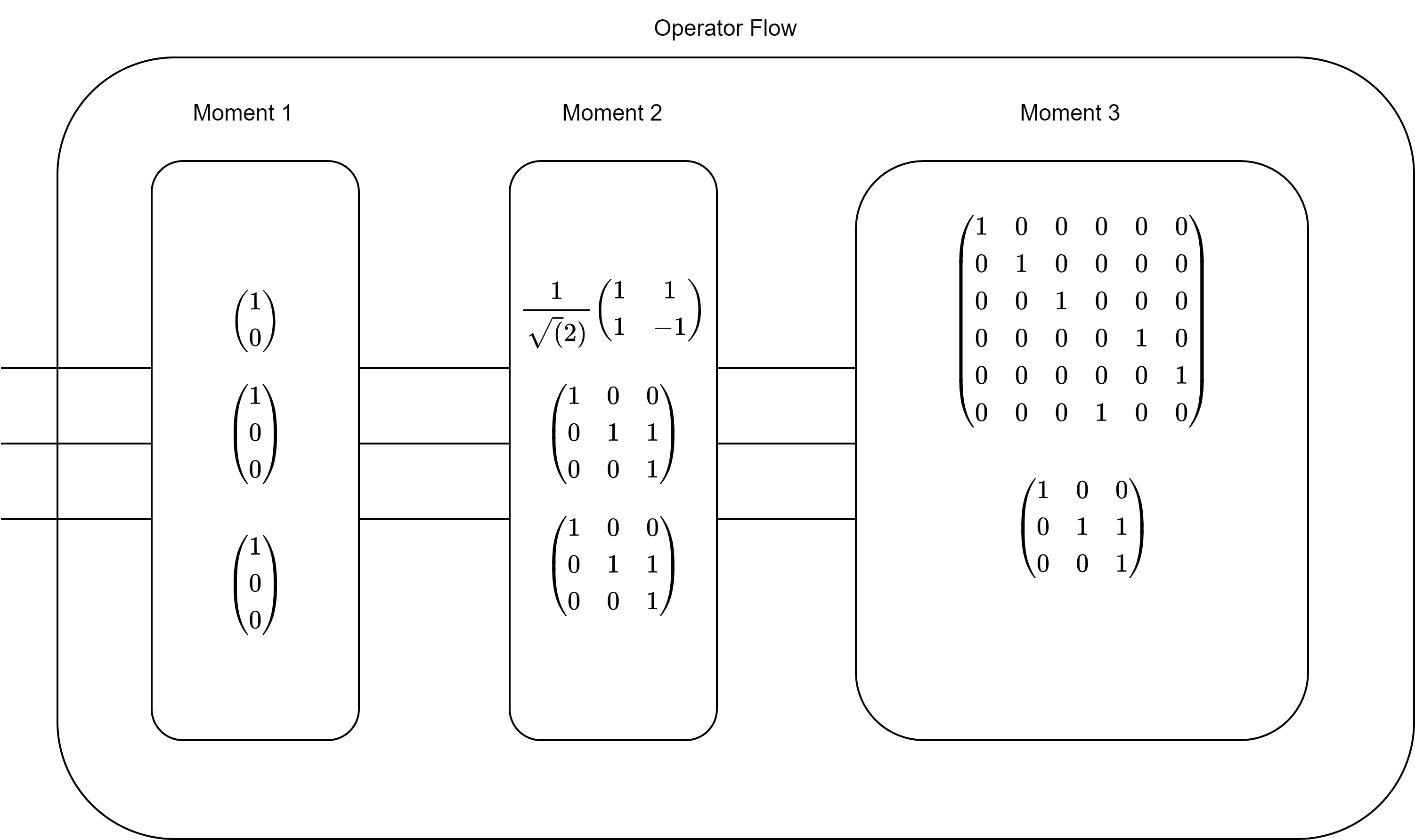}
\caption{The \texttt{OperatorFlow} and its internals before the circuit is compiled}
\label{exampleoperatorflow}
\end{figure}

The state of the \texttt{OperatorFlow} object and its internals are given as follows in Figure \ref{exampleoperatorflow}. 

Note that the tensor and inner products are not evaluated at the time of creation. In order to begin the process of execution, we invoke the following lines of code: \\
\texttt{qc.measure\_all()}\\
\texttt{qc.run()}

These will add measurement operators, interpreted here as flag variables, to signify the end of all operations within a circuit. The circuit is then executed with the \texttt{run()} command, which begins calculating the tensor and the inner products using the backend specified. The final output is as follows:\\
\\
\texttt{Build elapsed: 0.0001919269561767578s}\\
\texttt{Execution elapsed: 0.0010845661163330078s}\\
\\
\texttt{
    [
        \{'|000>': 0.7071067811865475\}, 
        \{'|120>': 0.7071067811865475\}
    ]
}

The final simulation result comes with loading-time and execution-time of the given circuit as shown in the above example. We also obtain the final output quantum states as $\ket{000}$ and $\ket{120}$ with amplitude 0.7071067811865475 for the example circuit.

\section{Experiments and Discussion}\label{4}

Apart from python console, \textbf{QuDiet} offers another form of circuit specification $i.e.,$ QuDiet's QASM. With the increasing usage of quantum circuit description, QASM (Quantum Assembly Language)~\cite{svoretoward} was introduced. Through QuDiet's QASM, one can declare the qubits or qudits and can describe the operations (gates) on those qubits or qudits to be run on \textbf{QuDiet}. For ease of understanding, a sample QASM program on \textbf{QuDiet} is presented as following:\\
\texttt{.qudit 3}\\
\texttt{qudit x0 (2)}\\
\texttt{qudit x1 (3)}\\
\texttt{qudit x2 (3)}\\
\texttt{.begin}\\
\texttt{X x0}\\
\texttt{H x0}\\
\texttt{Z x0}\\
\texttt{X x1}\\
\texttt{X X2 2}\\
\texttt{CX x0 x1}\\
\texttt{CX x1 x1 2}\\
\texttt{.end}\\
In this QASM program, we declare 3 qudits, one is qubit (\texttt{x0}), one is qutrit (\texttt{x1}) and last one is qutrit (\texttt{x2}). NOT, Hadamard and phase gate are applied on qubit \texttt{x0}. Then a generalized NOT gate with $+1$ is applied on qutrit \texttt{x1} followed by a generalized NOT gate with $+2$ is applied on qutrit \texttt{x2}. A generalized CNOT with $+1$ is on \texttt{x0} and \texttt{x1} and a generalized CNOT with $+2$ is on \texttt{x1} and \texttt{x2}. The most widely used QASM, $i.e.,$ OpenQASM \cite{openqasm} can be converted to the QuDiet's QASM form with the help of inbuilt lexer that is available in \textbf{QuDiet} to make it more user-friendly.

\paragraph{Benchmarking with state-of-the-art simulators}
We have taken 21 benchmark circuits as an initial benchmarking, ranging from 3 qubit-qutrit to 7 qubit-qutrit in the form of QASM from \cite{revlib} to verify our proposed simulator. To simulate all the 21 circuits, Toffoli gate is decomposed with intermediate qutrits as discussed earlier to get the algorithmic advantage. The simulation results are shown in Table \ref{Tab:simu}. The complete simulation time is based on three different parameters, (i) prepossessing-time; (ii) loading-time; and (iii) execution-time.  We run these circuits with two backends, Numpy and Sparse. The maximum run-time (loading-time + execution-time) of these circuits is 0.3 seconds, which is akin to \cite{cirq_developers_2022_6599601}, albeit the total simulation-time is much lower since the prepossessing-time is much higher for \cite{cirq_developers_2022_6599601} as gates are needed to be defined manually based on dimensions. The exact prepossessing-time of \cite{cirq_developers_2022_6599601} can never be determined since it is manual and very complicated to be defined mathematically. In our case, \textbf{QuDiet} being fully automatic, the prepossessing-time is negligible.   


\begin{table}[!ht]
    \caption{Loading and Execution Times \textit{(in milliseconds)} of benchmark circuits on the Numpy and Sparse Backend}
    \centering
    \begin{tabular}{|p{2.5cm}|p{1.5cm}|p{1.5cm}|p{1.5cm}|p{1.5cm}|p{1.5cm}|}
 \hline
  \multirow{2}{1.9cm}{circuit} & \multirow{2}{.8cm}{(width, depth)} & \multicolumn{2}{|c|}{Numpy Backend} &  \multicolumn{2}{|c|}{Sparse Backend}\\
 \cline{3-6}
          &   & loading-time & execution-time & loading-time & execution-time \\ \hline\hline
         toffoli\_2\_tof & (3, 3) & 0.289 & 1.219 & 1.193 & 3.062\\ \hline
         ex-1\_166\_tof & (3, 6) & 0.347 & 2.254 & 1.21 & 5.753\\ \hline
         3\_17\_14\_tof & (3, 10) & 0.937 & 4.43 & 3.33 & 14.656 \\ \hline
         3\_17\_13\_tof & (3, 10) & 1.363 & 33.272 & 3.33 & 14.656\\ \hline
         miller\_11\_tof & (3, 11) & 0.571 & 5.379 & 1.154 & 10.428\\ \hline
         decod24-v0\_38\_tof & (4, 12) & 1.055 & 8.435 & 1.509 & 16.738\\ \hline
         4\_49\_17\_tof & (4, 22) & 0.732 & 32.765 & 2.324 & 11.9348\\ \hline
         mod5d1\_63\_tof & (5, 9) & 0.435 & 11.948 & 1.339 & 17.745 \\ \hline
         mod5mils\_65\_tof & (5, 9) & 0.463 & 7.46 & 1.86 & 15.955\\ \hline
         4gt11\_82\_tof & (5, 14) & 0.579 & 17.099 & 1.458 & 26.949\\ \hline
         4mod5-v0\_18\_tof & (5, 16) & 0.68 & 27.89 & 2.505 & 34.822\\ \hline
         rd32\_270\_tof & (5, 17) & 0.78 & 22.968 & 2.1836 & 36.203\\ \hline
         alu-v0\_26\_tof & (5, 19) & 0.732 & 64.351 & 1.538 & 69.933\\ \hline
         4gt5\_76\_tof & (5, 26) & 0.951 & 41.93 & 2.311 & 54.336\\ \hline
         aj-e11\_165\_tof & (5, 33) & 1.076 & 103.887 & 2.531 & 102.521 \\ \hline
         4\_49\_16\_tof & (5, 48) & 1.675 & 125.8 & 2.324 & 119.348\\ \hline
         decod24-enable\_125\_tof & (6, 15) & 0.61 & 69.66 & 2.274 & 66.186\\ \hline
         decod24-bdd\_294\_tof & (6, 17) & 0.734 & 38.439 & 1.957 & 52.381\\ \hline
         4gt4-v0\_72\_tof & (6, 49)  & 1.577 & 315.283 & 2.555 & 129.431\\ \hline
         alu-bdd\_288\_tof & (7, 18) & 0.755 & 93.461 & 2.592 & 92.534\\ \hline
         4mod5-bdd\_287\_tof & (7, 22) & 0.701 & 112.74 & 2.61 & 88.45 \\ \hline
    \end{tabular}
    \label{Tab:simu}
\end{table}

We further simulate more larger circuits on our proposed simulator. The results are shown in Table \ref{sparsecuda}. These 17 medium-sized circuits are also taken from \cite{revlib}. It is exhibited through numerical simulation that these circuits are well executable with Sparse-cuda backend due to its dense nature as compared to other backends. It can also be noted that if these 38 qubit-only circuits without decomposition from Table \ref{Tab:simu} and \ref{sparsecuda} are simulated on \textbf{QuDiet}, the performence time is same as \cite{cirq_developers_2022_6599601, Qiskit}.

\begin{table}[!ht]
    \caption{Loading and Execution Times \textit{(in milliseconds)} of some benchmark circuits on Sparse-cuda Backend}
    \centering
    \begin{tabular}{|l|l|l|l|}
    \hline
    circuit & (width, depth) & loading-time & execution-time \\ \hline
    
hwb4\_49\_tof & (5, 51) & 2.58 & 760.01\\ \hline    
one-two-three-v0\_97\_tof & (5,  56) & 575.1 & 963.46\\ \hline
    
mod8-10\_177\_tof       & (6, 89) & 568.71 & 1656.48\\ \hline

mod5adder\_127\_tof & (6, 106) & 462.74 & 696.1\\ \hline

sf\_274\_tof & (6, 148) & 5.91 & 1717.76\\ \hline

ham7\_104\_tof & (7, 73) & 3.18 & 2196.92\\ \hline

C17\_204\_tof & (7, 106) & 4.4 & 5135.62\\ \hline

majority\_239\_tof & (7, 149) & 5.9 & 5966.78\\ \hline

sym6\_145\_tof &        (7, 945) &  24.09 & 43371.93\\ \hline

f2\_232\_tof & (8, 293) & 9.59 & 19534.77\\ \hline

con1\_216\_tof & (9, 227) & 10.2 & 25159.1\\ \hline

mini\_alu\_305\_tof & (10, 39) & 784.24 & 884254.49\\ \hline

sys6-v0\_111\_tof & (10, 46) & 7.4 & 2060.27\\ \hline

wim\_266\_tof & (11, 219) & 7.87 & 997068.12\\ \hline

dc1\_220\_tof & (11, 435) & 13.96 & 2329661.45\\ \hline

0410184\_169\_tof & (14, 60) & 6.58 &  337.46\\ \hline
     
    \end{tabular}
    \label{sparsecuda}
\end{table}

\begin{figure}[h]
    \centering
    \includegraphics[scale=0.23]{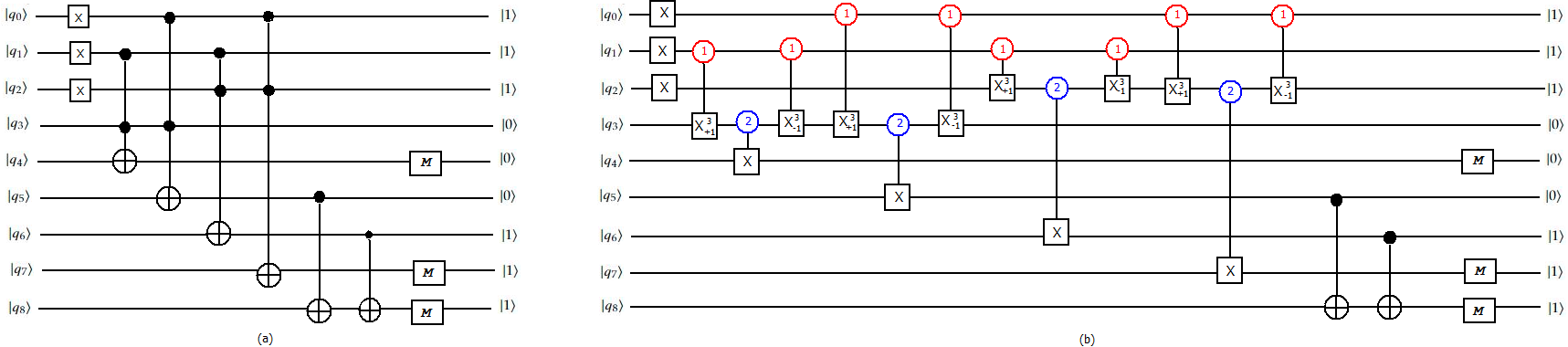}
    \caption{(a) Quantum Multiplier Circuit for the multiplication of $3 \times 2$; (b) Quantum Multiplier with Intermediate Qutrit for the multiplication of $3 \times 2$.}
    \label{fig:multiplier}
\end{figure}

 We have employed the multiplication of $3 \times 2$ as an example to illustrate our simulator's efficiency in designing a quantum multiplier with intermediate qutrit. In Figure \ref{fig:multiplier}(a), in light of the preceding example, a multiplier circuit has been provided in accordance with \cite{revlib}, in which all the qubits are initialized with $\ket{0}$. In this circuit, the first four qubits ($q_0$ - $q_3$) are the input qubits, where the first two qubits ($q_0$ and $q_1$) represent the number 3 by applying two NOT gates on them and the other two qubits ($q_2$ and $q_3$) represent the number 2 by applying NOT gate on qubit $q_2$. Subsequently, using Toffoli gates, we conduct a multiply operation on these qubits and store the result in ancilla qubits ($q_4$ - $q_{7}$). Now, using CNOT gates on an ancilla qubit $q_8$, we execute addition. Lastly, to obtain the resultant output of $3 \times 2$, we need to measure the qubits ($q_4$, $q_{7}$ and $q_{8}$). Additionally, each of the Toffoli gates shown in Fig. \ref{fig:multiplier}(a) are realized with the help of the intermediate qutrit method as shown in Fig. \ref{fig:multiplier}(b) to achieve asymptotic advancement of the circuit. Our numerical simulation on \textbf{QuDiet} also yields $3 \times 2 = 6$ appropriately. Our simulation results further show that if we use Numpy backend, then the loading-time is 34.631 (ms) and execution-time is 136.532 (ms). For Sparse beckend, the loading-time is 17.157 (ms) and the execution-time is 1.865 (s) for the multiplier of $3 \times 2$ circuit, whose width is 9 and depth is 15. 

\subsection{Simulation of Some Well-known Quantum Algorithms}

We simulate some well-known quantum algorithms like, Grover's algorithm, Simon algorithm, Berstein-Vazirani algorithm etc on \textbf{QuDiet} considering different backends and amalgamate them as a package for faster prototyping. The result of the simulation is shown in Table \ref{algo}. The trade-off of loading-time and execution-time between different backends for various quantum algorithms is exhibited in Table \ref{algo}. As a concluding remark, using the \textbf{QuDiet} simulator, we also simulate a comparatively larger circuit of 20 qutrits on depth 80 quite comprehensibly. The simulation has been performed on a local computer with processor Intel(R) Core(TM) i5-6300U CPU \@ 2.40 GHz 2.50 GHz, RAM 8.00 GB, and 64-bit windows operating system.

\begin{table}[!ht]
    \caption{Loading and Execution Times \textit{(in milliseconds)} of some supremacy circuits on various Backend}
    \centering
    \begin{tabular}{|l|l|l|l|}
    \hline
    circuit (backend) & (width, depth) & loading-time  & execution-time \\ \hline
       Grover\_n2 (sparse) & (2, 9) & 2.55 & 25.76 \\ \hline
    Grover\_n2 (cuda) & (2, 9) & 1.16 & 248.16 \\ \hline
   Grover\_n2 (sparse-cuda) & (2, 9) & 1.19 & 44.05 \\ \hline
    Simon\_n6 (sparse) & (6, 13) & 5.39 & 92.94 \\ \hline
 Simon\_n6 (cuda) & (6, 13) & 1.05 & 30.7 \\ \hline
    Simon\_n6 (sparse-cuda) & (6, 13) & 2.11 & 183.22 \\ \hline
    Seca\_n11 (cuda) & (11, 69) & 180.8 & 13536.19 \\ \hline
   Seca\_n11 (sparse-cuda) & (11, 69) & 524.42 & 12008.1 \\ \hline
 SAT\_n11 (cuda) & (11, 139) & 5.9 & 16861.99 \\ \hline
 SAT\_n11 (sparse-cuda) & (11, 139) & 8.62 & 12229.81 \\ \hline
 BV\_n14 (cuda) & (14, 16) & 2.25 & 75112.75 \\ \hline
     BV\_n14 (sparse-cuda) & (14, 16) & 4.65 & 932159.35 \\ \hline
    \end{tabular}
    \label{algo}
\end{table}


\section{Future Scope}\label{5}
In this section, we provide a glimpse of the future version of \textbf{QuDiet}. In future, we can use a multi-threaded C++ core
that uses AVX instructions, which can achieve state-of-the-art
performance in matrix-vector multiplication of quantum
states. The C++ core, which uses a similar syntax
of numpy.dot can be integrated into our simulator whenever
a matrix-vector multiplication is needed.
Formation of a DAG interfaces for better optimization workflows can be incorporated in the future.
 Large-scale simulations on heterogeneous HPC clusters via MPI can also be looked into. Memorization and lookup tables for fast simulation can be a good case-study for future version of \textbf{QuDiet}. Simulation with noise models shall give a more broader picture of qudit simulation. 
Interfaces for different quantum hardware topologies shall need to be taken care in the future version of \textbf{QuDiet} to reduce the gap between physical and logical qudit simulation.

\section{Conclusion}\label{6}

In this paper, we introduced \textbf{QuDiet}, a hybrid qubit-qudit simulator that can classically simulate any finite-dimensional quantum system. It is exhibited that the \textbf{QuDiet} offers user-friendly environment to simulate qudit systems with an abstraction.  \textbf{QuDiet} is efficient since generalized gates can be easily used without spending much time to defining them manually during simulation. We also showed considerable speed-up in simulation time for benchmark circuits. Finally, we simulated some well-known quantum algorithms in qudit setting to analysis the performance of our proposed simulator  \textbf{QuDiet}. Furthermore, other available platforms can integrate \textbf{QuDiet} as a classical simulation option for higher-dimensional quantum systems to their platforms, since \textbf{QuDiet} is an open-source python-based simulator.

\bibliographystyle{ACM-Reference-Format}
\bibliography{sample-base}


\begin{thebibliography}{36}


\ifx \showCODEN    \undefined \def \showCODEN     #1{\unskip}     \fi
\ifx \showDOI      \undefined \def \showDOI       #1{#1}\fi
\ifx \showISBNx    \undefined \def \showISBNx     #1{\unskip}     \fi
\ifx \showISBNxiii \undefined \def \showISBNxiii  #1{\unskip}     \fi
\ifx \showISSN     \undefined \def \showISSN      #1{\unskip}     \fi
\ifx \showLCCN     \undefined \def \showLCCN      #1{\unskip}     \fi
\ifx \shownote     \undefined \def \shownote      #1{#1}          \fi
\ifx \showarticletitle \undefined \def \showarticletitle #1{#1}   \fi
\ifx \showURL      \undefined \def \showURL       {\relax}        \fi
\providecommand\bibfield[2]{#2}
\providecommand\bibinfo[2]{#2}
\providecommand\natexlab[1]{#1}
\providecommand\showeprint[2][]{arXiv:#2}

\bibitem[\protect\citeauthoryear{Adcock, Høyer, and Sanders}{Adcock
  et~al\mbox{.}}{2016}]%
        {Adcock_2016}
\bibfield{author}{\bibinfo{person}{M.~R.~A. Adcock}, \bibinfo{person}{P.
  Høyer}, {and} \bibinfo{person}{B.~C. Sanders}.}
  \bibinfo{year}{2016}\natexlab{}.
\newblock \showarticletitle{Quantum computation with coherent spin states and
  the close Hadamard problem}.
\newblock \bibinfo{journal}{\emph{Quantum Information Processing}}
  \bibinfo{volume}{15}, \bibinfo{number}{4} (\bibinfo{date}{Jan}
  \bibinfo{year}{2016}), \bibinfo{pages}{1361–1386}.
\newblock
\showISSN{1573-1332}
\urldef\tempurl%
\url{https://doi.org/10.1007/s11128-015-1229-0}
\showDOI{\tempurl}


\bibitem[\protect\citeauthoryear{Barenco, Bennett, Cleve, DiVincenzo, Margolus,
  Shor, Sleator, Smolin, and Weinfurter}{Barenco et~al\mbox{.}}{1995}]%
        {barenco}
\bibfield{author}{\bibinfo{person}{A. Barenco}, \bibinfo{person}{C.~H.
  Bennett}, \bibinfo{person}{R. Cleve}, \bibinfo{person}{D.~P. DiVincenzo},
  \bibinfo{person}{N. Margolus}, \bibinfo{person}{P. Shor}, \bibinfo{person}{T.
  Sleator}, \bibinfo{person}{J.~A. Smolin}, {and} \bibinfo{person}{H.
  Weinfurter}.} \bibinfo{year}{1995}\natexlab{}.
\newblock \showarticletitle{Elementary gates for quantum computation}.
\newblock \bibinfo{journal}{\emph{Physical Review A}}  \bibinfo{volume}{52}
  (\bibinfo{date}{Nov} \bibinfo{year}{1995}), \bibinfo{pages}{3457--3467}.
\newblock
Issue 5.
\urldef\tempurl%
\url{https://doi.org/10.1103/PhysRevA.52.3457}
\showDOI{\tempurl}


\bibitem[\protect\citeauthoryear{Bartlett, de~Guise, and Sanders}{Bartlett
  et~al\mbox{.}}{2002}]%
        {Bartlett_2002}
\bibfield{author}{\bibinfo{person}{S.~D. Bartlett}, \bibinfo{person}{H. de
  Guise}, {and} \bibinfo{person}{B.~C. Sanders}.}
  \bibinfo{year}{2002}\natexlab{}.
\newblock \showarticletitle{Quantum encodings in spin systems and harmonic
  oscillators}.
\newblock \bibinfo{journal}{\emph{Physical Review A}} \bibinfo{volume}{65},
  \bibinfo{number}{5} (\bibinfo{date}{May} \bibinfo{year}{2002}).
\newblock
\showISSN{1094-1622}
\urldef\tempurl%
\url{https://doi.org/10.1103/physreva.65.052316}
\showDOI{\tempurl}


\bibitem[\protect\citeauthoryear{Bello, Challenger, Cross, Faro, Gambetta,
  Gomez, Abhari, Martin, Moreda, Perez, Winston, and Wood}{Bello
  et~al\mbox{.}}{2021}]%
        {Qiskit}
\bibfield{author}{\bibinfo{person}{Luciano Bello}, \bibinfo{person}{Jim
  Challenger}, \bibinfo{person}{Andrew Cross}, \bibinfo{person}{Ismael Faro},
  \bibinfo{person}{Jay Gambetta}, \bibinfo{person}{Juan Gomez},
  \bibinfo{person}{Ali~J. Abhari}, \bibinfo{person}{Paco Martin},
  \bibinfo{person}{Diego Moreda}, \bibinfo{person}{Jesus Perez},
  \bibinfo{person}{Erick Winston}, {and} \bibinfo{person}{Chris Wood}.}
  \bibinfo{year}{2021}\natexlab{}.
\newblock \bibinfo{title}{Qiskit: An Open-source Framework for Quantum
  Computing}.
\newblock
\newblock
\urldef\tempurl%
\url{https://doi.org/10.5281/zenodo.2573505}
\showDOI{\tempurl}


\bibitem[\protect\citeauthoryear{Bocharov, Cui, Roetteler, and Svore}{Bocharov
  et~al\mbox{.}}{2015}]%
        {bocharov2015improved}
\bibfield{author}{\bibinfo{person}{Alex Bocharov}, \bibinfo{person}{Shawn~X.
  Cui}, \bibinfo{person}{Martin Roetteler}, {and} \bibinfo{person}{Krysta~M.
  Svore}.} \bibinfo{year}{2015}\natexlab{}.
\newblock \bibinfo{title}{Improved Quantum Ternary Arithmetics}.
\newblock
\newblock
\showeprint[arxiv]{1512.03824}~[quant-ph]


\bibitem[\protect\citeauthoryear{Bocharov, Roetteler, and Svore}{Bocharov
  et~al\mbox{.}}{2017}]%
        {Bocharov_2017}
\bibfield{author}{\bibinfo{person}{Alex Bocharov}, \bibinfo{person}{Martin
  Roetteler}, {and} \bibinfo{person}{Krysta~M. Svore}.}
  \bibinfo{year}{2017}\natexlab{}.
\newblock \showarticletitle{Factoring with qutrits: Shor’s algorithm on
  ternary and metaplectic quantum architectures}.
\newblock \bibinfo{journal}{\emph{Physical Review A}} \bibinfo{volume}{96},
  \bibinfo{number}{1} (\bibinfo{date}{Jul} \bibinfo{year}{2017}).
\newblock
\showISSN{2469-9934}
\urldef\tempurl%
\url{https://doi.org/10.1103/physreva.96.012306}
\showDOI{\tempurl}


\bibitem[\protect\citeauthoryear{Cao, Peng, Zheng, and Long}{Cao
  et~al\mbox{.}}{2011}]%
        {qft}
\bibfield{author}{\bibinfo{person}{Ye Cao}, \bibinfo{person}{Shi-Guo Peng},
  \bibinfo{person}{Chao Zheng}, {and} \bibinfo{person}{Gui Long}.}
  \bibinfo{year}{2011}\natexlab{}.
\newblock \showarticletitle{Quantum Fourier Transform and Phase Estimation in
  Qudit System}.
\newblock \bibinfo{journal}{\emph{Communications in Theoretical Physics}}
  \bibinfo{volume}{55} (\bibinfo{date}{05} \bibinfo{year}{2011}),
  \bibinfo{pages}{790--794}.
\newblock
\urldef\tempurl%
\url{https://doi.org/10.1088/0253-6102/55/5/11}
\showDOI{\tempurl}


\bibitem[\protect\citeauthoryear{Cross, Bishop, Smolin, and Gambetta}{Cross
  et~al\mbox{.}}{2017}]%
        {openqasm}
\bibfield{author}{\bibinfo{person}{Andrew~W. Cross}, \bibinfo{person}{Lev~S.
  Bishop}, \bibinfo{person}{John~A. Smolin}, {and} \bibinfo{person}{Jay~M.
  Gambetta}.} \bibinfo{year}{2017}\natexlab{}.
\newblock \bibinfo{title}{Open Quantum Assembly Language}.
\newblock
\newblock
\urldef\tempurl%
\url{https://doi.org/10.48550/ARXIV.1707.03429}
\showDOI{\tempurl}


\bibitem[\protect\citeauthoryear{Cui, Hong, and Wang}{Cui
  et~al\mbox{.}}{2015}]%
        {Cui_2015}
\bibfield{author}{\bibinfo{person}{S.~X. Cui}, \bibinfo{person}{S.\-M. Hong},
  {and} \bibinfo{person}{Z. Wang}.} \bibinfo{year}{2015}\natexlab{}.
\newblock \showarticletitle{Universal quantum computation with weakly integral
  anyons}.
\newblock \bibinfo{journal}{\emph{Quantum Information Processing}}
  \bibinfo{volume}{14}, \bibinfo{number}{8} (\bibinfo{date}{May}
  \bibinfo{year}{2015}), \bibinfo{pages}{2687–2727}.
\newblock
\showISSN{1573-1332}
\urldef\tempurl%
\url{https://doi.org/10.1007/s11128-015-1016-y}
\showDOI{\tempurl}


\bibitem[\protect\citeauthoryear{Cui and Wang}{Cui and Wang}{2015}]%
        {Cui_2015first}
\bibfield{author}{\bibinfo{person}{S.~X. Cui} {and} \bibinfo{person}{Z. Wang}.}
  \bibinfo{year}{2015}\natexlab{}.
\newblock \showarticletitle{Universal quantum computation with metaplectic
  anyons}.
\newblock \bibinfo{journal}{\emph{J. Math. Phys.}} \bibinfo{volume}{56},
  \bibinfo{number}{3} (\bibinfo{date}{Mar} \bibinfo{year}{2015}),
  \bibinfo{pages}{032202}.
\newblock
\showISSN{1089-7658}
\urldef\tempurl%
\url{https://doi.org/10.1063/1.4914941}
\showDOI{\tempurl}


\bibitem[\protect\citeauthoryear{Developers}{Developers}{2022}]%
        {cirq_developers_2022_6599601}
\bibfield{author}{\bibinfo{person}{Cirq Developers}.}
  \bibinfo{year}{2022}\natexlab{}.
\newblock \bibinfo{booktitle}{\emph{Cirq}}.
\newblock
\urldef\tempurl%
\url{https://doi.org/10.5281/zenodo.6599601}
\showDOI{\tempurl}
\newblock
\shownote{{See full list of authors on Github: https://github
  .com/quantumlib/Cirq/graphs/contributors}.}


\bibitem[\protect\citeauthoryear{Di and Wei}{Di and Wei}{2013}]%
        {Di_2013}
\bibfield{author}{\bibinfo{person}{Yao-Min Di} {and} \bibinfo{person}{Hai~Rui
  Wei}.} \bibinfo{year}{2013}\natexlab{}.
\newblock \showarticletitle{Synthesis of multivalued quantum logic circuits by
  elementary gates}.
\newblock \bibinfo{journal}{\emph{Physical Review A}} \bibinfo{volume}{87},
  \bibinfo{number}{1} (\bibinfo{date}{Jan} \bibinfo{year}{2013}).
\newblock
\showISSN{1094-1622}
\urldef\tempurl%
\url{https://doi.org/10.1103/physreva.87.012325}
\showDOI{\tempurl}


\bibitem[\protect\citeauthoryear{Dogra, Arvind, and Dorai}{Dogra
  et~al\mbox{.}}{2014}]%
        {Dogra_2014}
\bibfield{author}{\bibinfo{person}{S. Dogra}, \bibinfo{person}{Arvind}, {and}
  \bibinfo{person}{K. Dorai}.} \bibinfo{year}{2014}\natexlab{}.
\newblock \showarticletitle{Determining the parity of a permutation using an
  experimental NMR qutrit}.
\newblock \bibinfo{journal}{\emph{Physics Letters A}} \bibinfo{volume}{378},
  \bibinfo{number}{46} (\bibinfo{date}{Oct} \bibinfo{year}{2014}),
  \bibinfo{pages}{3452–3456}.
\newblock
\showISSN{0375-9601}
\urldef\tempurl%
\url{https://doi.org/10.1016/j.physleta.2014.10.003}
\showDOI{\tempurl}


\bibitem[\protect\citeauthoryear{Fan}{Fan}{2007}]%
        {Fan_2007}
\bibfield{author}{\bibinfo{person}{Y. Fan}.} \bibinfo{year}{2007}\natexlab{}.
\newblock \showarticletitle{A Generalization of the Deutsch-Jozsa Algorithm to
  Multi-Valued Quantum Logic}. In \bibinfo{booktitle}{\emph{37th International
  Symposium on Multiple-Valued Logic (ISMVL'07)}}. \bibinfo{publisher}{IEEE
  Computer Society}, \bibinfo{address}{Los Alamitos, CA, USA},
  \bibinfo{pages}{12}.
\newblock
\showISSN{0195-623X}
\urldef\tempurl%
\url{https://doi.org/10.1109/ISMVL.2007.3}
\showDOI{\tempurl}


\bibitem[\protect\citeauthoryear{Farhi and Gutmann}{Farhi and Gutmann}{1998}]%
        {Farhi_1998}
\bibfield{author}{\bibinfo{person}{Edward Farhi} {and} \bibinfo{person}{Sam
  Gutmann}.} \bibinfo{year}{1998}\natexlab{}.
\newblock \showarticletitle{Quantum computation and decision trees}.
\newblock \bibinfo{journal}{\emph{Physical Review A}} \bibinfo{volume}{58},
  \bibinfo{number}{2} (\bibinfo{date}{Aug} \bibinfo{year}{1998}),
  \bibinfo{pages}{915–928}.
\newblock
\showISSN{1094-1622}
\urldef\tempurl%
\url{https://doi.org/10.1103/physreva.58.915}
\showDOI{\tempurl}


\bibitem[\protect\citeauthoryear{Gao, Ji, Tan, and Zhao}{Gao
  et~al\mbox{.}}{2020b}]%
        {sparse}
\bibfield{author}{\bibinfo{person}{Jianhua Gao}, \bibinfo{person}{Weixing Ji},
  \bibinfo{person}{Zhaonian Tan}, {and} \bibinfo{person}{Yueyan Zhao}.}
  \bibinfo{year}{2020}\natexlab{b}.
\newblock \bibinfo{title}{A Systematic Survey of General Sparse Matrix-Matrix
  Multiplication}.
\newblock
\newblock
\urldef\tempurl%
\url{https://doi.org/10.48550/ARXIV.2002.11273}
\showDOI{\tempurl}


\bibitem[\protect\citeauthoryear{Gao, Erhard, Zeilinger, and Krenn}{Gao
  et~al\mbox{.}}{2020a}]%
        {Gao_2020}
\bibfield{author}{\bibinfo{person}{X. Gao}, \bibinfo{person}{M. Erhard},
  \bibinfo{person}{A. Zeilinger}, {and} \bibinfo{person}{M. Krenn}.}
  \bibinfo{year}{2020}\natexlab{a}.
\newblock \showarticletitle{Computer-Inspired Concept for High-Dimensional
  Multipartite Quantum Gates}.
\newblock \bibinfo{journal}{\emph{Physical Review Letters}}
  \bibinfo{volume}{125}, \bibinfo{number}{5} (\bibinfo{date}{Jul}
  \bibinfo{year}{2020}).
\newblock
\showISSN{1079-7114}
\urldef\tempurl%
\url{https://doi.org/10.1103/physrevlett.125.050501}
\showDOI{\tempurl}


\bibitem[\protect\citeauthoryear{Gedik, Silva, Çakmak, Karpat, Vidoto,
  Soares-Pinto, deAzevedo, and Fanchini}{Gedik et~al\mbox{.}}{2015}]%
        {Gedik_2015}
\bibfield{author}{\bibinfo{person}{Z. Gedik}, \bibinfo{person}{I.~A. Silva},
  \bibinfo{person}{B. Çakmak}, \bibinfo{person}{G. Karpat},
  \bibinfo{person}{E.~L.~G. Vidoto}, \bibinfo{person}{D.~O. Soares-Pinto},
  \bibinfo{person}{E.~R. deAzevedo}, {and} \bibinfo{person}{F.~F. Fanchini}.}
  \bibinfo{year}{2015}\natexlab{}.
\newblock \showarticletitle{Computational speed-up with a single qudit}.
\newblock \bibinfo{journal}{\emph{Scientific Reports}} \bibinfo{volume}{5},
  \bibinfo{number}{1} (\bibinfo{date}{Oct} \bibinfo{year}{2015}).
\newblock
\showISSN{2045-2322}
\urldef\tempurl%
\url{https://doi.org/10.1038/srep14671}
\showDOI{\tempurl}


\bibitem[\protect\citeauthoryear{Giraldo-Carvajal, Duque-Ramirez, and
  Jaramillo-Villegas}{Giraldo-Carvajal et~al\mbox{.}}{2021}]%
        {sky}
\bibfield{author}{\bibinfo{person}{Andres Giraldo-Carvajal},
  \bibinfo{person}{Daniel~A. Duque-Ramirez}, {and} \bibinfo{person}{Jose~A.
  Jaramillo-Villegas}.} \bibinfo{year}{2021}\natexlab{}.
\newblock \bibinfo{title}{QuantumSkynet: A High-Dimensional Quantum Computing
  Simulator}.
\newblock
\newblock
\urldef\tempurl%
\url{https://doi.org/10.48550/ARXIV.2106.15833}
\showDOI{\tempurl}


\bibitem[\protect\citeauthoryear{Gokhale, Baker, Duckering, Brown, Brown, and
  Chong}{Gokhale et~al\mbox{.}}{2019}]%
        {Gokhale_2019}
\bibfield{author}{\bibinfo{person}{Pranav Gokhale},
  \bibinfo{person}{Jonathan~M. Baker}, \bibinfo{person}{Casey Duckering},
  \bibinfo{person}{Natalie~C. Brown}, \bibinfo{person}{Kenneth~R. Brown}, {and}
  \bibinfo{person}{Frederic~T. Chong}.} \bibinfo{year}{2019}\natexlab{}.
\newblock \showarticletitle{Asymptotic improvements to quantum circuits via
  qutrits}.
\newblock \bibinfo{journal}{\emph{Proceedings of the 46th International
  Symposium on Computer Architecture}} (\bibinfo{date}{Jun}
  \bibinfo{year}{2019}).
\newblock
\showISBNx{9781450366694}
\urldef\tempurl%
\url{https://doi.org/10.1145/3307650.3322253}
\showDOI{\tempurl}


\bibitem[\protect\citeauthoryear{Khan and Perkowski}{Khan and
  Perkowski}{2006}]%
        {Khan_2006}
\bibfield{author}{\bibinfo{person}{F.~S. Khan} {and} \bibinfo{person}{M.
  Perkowski}.} \bibinfo{year}{2006}\natexlab{}.
\newblock \showarticletitle{Synthesis of multi-qudit hybrid and d-valued
  quantum logic circuits by decomposition}.
\newblock \bibinfo{journal}{\emph{Theoretical Computer Science}}
  \bibinfo{volume}{367}, \bibinfo{number}{3} (\bibinfo{date}{Dec}
  \bibinfo{year}{2006}), \bibinfo{pages}{336–346}.
\newblock
\showISSN{0304-3975}
\urldef\tempurl%
\url{https://doi.org/10.1016/j.tcs.2006.09.006}
\showDOI{\tempurl}


\bibitem[\protect\citeauthoryear{Klimov, Guzm\'an, Retamal, and
  Saavedra}{Klimov et~al\mbox{.}}{2003}]%
        {qutrit}
\bibfield{author}{\bibinfo{person}{A.~B. Klimov}, \bibinfo{person}{R.
  Guzm\'an}, \bibinfo{person}{J.~C. Retamal}, {and} \bibinfo{person}{C.
  Saavedra}.} \bibinfo{year}{2003}\natexlab{}.
\newblock \showarticletitle{Qutrit quantum computer with trapped ions}.
\newblock \bibinfo{journal}{\emph{Physical Review A}}  \bibinfo{volume}{67}
  (\bibinfo{date}{Jun} \bibinfo{year}{2003}), \bibinfo{pages}{062313}.
\newblock
Issue 6.
\urldef\tempurl%
\url{https://doi.org/10.1103/PhysRevA.67.062313}
\showDOI{\tempurl}


\bibitem[\protect\citeauthoryear{Koch, Yu, Gambetta, Houck, Schuster, Majer,
  Blais, Devoret, Girvin, and Schoelkopf}{Koch et~al\mbox{.}}{2007}]%
        {PhysRevA.76.042319}
\bibfield{author}{\bibinfo{person}{J. Koch}, \bibinfo{person}{T.~M. Yu},
  \bibinfo{person}{J. Gambetta}, \bibinfo{person}{A.~A. Houck},
  \bibinfo{person}{D.~I. Schuster}, \bibinfo{person}{J. Majer},
  \bibinfo{person}{A. Blais}, \bibinfo{person}{M.~H. Devoret},
  \bibinfo{person}{S.~M. Girvin}, {and} \bibinfo{person}{R.~J. Schoelkopf}.}
  \bibinfo{year}{2007}\natexlab{}.
\newblock \showarticletitle{Charge-insensitive qubit design derived from the
  Cooper pair box}.
\newblock \bibinfo{journal}{\emph{Phys. Rev. A}}  \bibinfo{volume}{76}
  (\bibinfo{date}{Oct} \bibinfo{year}{2007}), \bibinfo{pages}{042319}.
\newblock
Issue 4.
\urldef\tempurl%
\url{https://doi.org/10.1103/PhysRevA.76.042319}
\showDOI{\tempurl}


\bibitem[\protect\citeauthoryear{LaRose}{LaRose}{2019}]%
        {LaRose_2019}
\bibfield{author}{\bibinfo{person}{Ryan LaRose}.}
  \bibinfo{year}{2019}\natexlab{}.
\newblock \showarticletitle{Overview and Comparison of Gate Level Quantum
  Software Platforms}.
\newblock \bibinfo{journal}{\emph{Quantum}}  \bibinfo{volume}{3}
  (\bibinfo{date}{mar} \bibinfo{year}{2019}), \bibinfo{pages}{130}.
\newblock
\urldef\tempurl%
\url{https://doi.org/10.22331/q-2019-03-25-130}
\showDOI{\tempurl}


\bibitem[\protect\citeauthoryear{Leuenberger and Loss}{Leuenberger and
  Loss}{2001}]%
        {Leuenberger_2001}
\bibfield{author}{\bibinfo{person}{M.~N. Leuenberger} {and} \bibinfo{person}{D.
  Loss}.} \bibinfo{year}{2001}\natexlab{}.
\newblock \showarticletitle{Quantum computing in molecular magnets}.
\newblock \bibinfo{journal}{\emph{Nature}} \bibinfo{volume}{410},
  \bibinfo{number}{6830} (\bibinfo{date}{Apr} \bibinfo{year}{2001}),
  \bibinfo{pages}{789–793}.
\newblock
\showISSN{1476-4687}
\urldef\tempurl%
\url{https://doi.org/10.1038/35071024}
\showDOI{\tempurl}


\bibitem[\protect\citeauthoryear{Muthukrishnan and Stroud}{Muthukrishnan and
  Stroud}{2000}]%
        {Muthukrishnan_2000}
\bibfield{author}{\bibinfo{person}{Ashok Muthukrishnan} {and}
  \bibinfo{person}{C.~R. Stroud}.} \bibinfo{year}{2000}\natexlab{}.
\newblock \showarticletitle{Multivalued logic gates for quantum computation}.
\newblock \bibinfo{journal}{\emph{Physical Review A}} \bibinfo{volume}{62},
  \bibinfo{number}{5} (\bibinfo{date}{Oct} \bibinfo{year}{2000}).
\newblock
\showISSN{1094-1622}
\urldef\tempurl%
\url{https://doi.org/10.1103/physreva.62.052309}
\showDOI{\tempurl}


\bibitem[\protect\citeauthoryear{Nielsen and Chuang}{Nielsen and
  Chuang}{2010}]%
        {nielsen_chuang_2010}
\bibfield{author}{\bibinfo{person}{Michael~A. Nielsen} {and}
  \bibinfo{person}{Isaac~L. Chuang}.} \bibinfo{year}{2010}\natexlab{}.
\newblock \bibinfo{booktitle}{\emph{Quantum Computation and Quantum
  Information: 10th Anniversary Edition}}.
\newblock \bibinfo{publisher}{Cambridge University Press}.
\newblock
\urldef\tempurl%
\url{https://doi.org/10.1017/CBO9780511976667}
\showDOI{\tempurl}


\bibitem[\protect\citeauthoryear{Preskill}{Preskill}{2018}]%
        {Preskill_2018}
\bibfield{author}{\bibinfo{person}{J. Preskill}.}
  \bibinfo{year}{2018}\natexlab{}.
\newblock \showarticletitle{Quantum Computing in the NISQ era and beyond}.
\newblock \bibinfo{journal}{\emph{Quantum}}  \bibinfo{volume}{2}
  (\bibinfo{date}{Aug} \bibinfo{year}{2018}), \bibinfo{pages}{79}.
\newblock
\showISSN{2521-327X}
\urldef\tempurl%
\url{https://doi.org/10.22331/q-2018-08-06-79}
\showDOI{\tempurl}


\bibitem[\protect\citeauthoryear{Saha, Majumdar, Saha, Chakrabarti, and
  Sur-Kolay}{Saha et~al\mbox{.}}{2022}]%
        {sahapra}
\bibfield{author}{\bibinfo{person}{Amit Saha}, \bibinfo{person}{Ritajit
  Majumdar}, \bibinfo{person}{Debasri Saha}, \bibinfo{person}{Amlan
  Chakrabarti}, {and} \bibinfo{person}{Susmita Sur-Kolay}.}
  \bibinfo{year}{2022}\natexlab{}.
\newblock \showarticletitle{Asymptotically improved circuit for a $d$-ary
  Grover's algorithm with advanced decomposition of the $n$-qudit Toffoli
  gate}.
\newblock \bibinfo{journal}{\emph{Phys. Rev. A}}  \bibinfo{volume}{105}
  (\bibinfo{date}{Jun} \bibinfo{year}{2022}), \bibinfo{pages}{062453}.
\newblock
Issue 6.
\urldef\tempurl%
\url{https://doi.org/10.1103/PhysRevA.105.062453}
\showDOI{\tempurl}


\bibitem[\protect\citeauthoryear{Saha, Mandal, Saha, and Chakrabarti}{Saha
  et~al\mbox{.}}{2021}]%
        {9410395}
\bibfield{author}{\bibinfo{person}{Amit Saha}, \bibinfo{person}{Sudhindu~Bikash
  Mandal}, \bibinfo{person}{Debasri Saha}, {and} \bibinfo{person}{Amlan
  Chakrabarti}.} \bibinfo{year}{2021}\natexlab{}.
\newblock \showarticletitle{One-Dimensional Lazy Quantum Walk in Ternary
  System}.
\newblock \bibinfo{journal}{\emph{IEEE Transactions on Quantum Engineering}}
  \bibinfo{volume}{2} (\bibinfo{year}{2021}), \bibinfo{pages}{1--12}.
\newblock
\urldef\tempurl%
\url{https://doi.org/10.1109/TQE.2021.3074707}
\showDOI{\tempurl}


\bibitem[\protect\citeauthoryear{Smith, Curtis, and Zeng}{Smith
  et~al\mbox{.}}{2017}]%
        {rigetti}
\bibfield{author}{\bibinfo{person}{Robert~S. Smith},
  \bibinfo{person}{Michael~J. Curtis}, {and} \bibinfo{person}{William~J.
  Zeng}.} \bibinfo{year}{2017}\natexlab{}.
\newblock \bibinfo{title}{A Practical Quantum Instruction Set Architecture}.
\newblock
\newblock
\showeprint[arxiv]{1608.03355}~[quant-ph]


\bibitem[\protect\citeauthoryear{Steiger, H{\"{a}}ner, and Troyer}{Steiger
  et~al\mbox{.}}{2018}]%
        {Steiger2018projectqopensource}
\bibfield{author}{\bibinfo{person}{Damian~S. Steiger}, \bibinfo{person}{Thomas
  H{\"{a}}ner}, {and} \bibinfo{person}{Matthias Troyer}.}
  \bibinfo{year}{2018}\natexlab{}.
\newblock \showarticletitle{Project{Q}: an open source software framework for
  quantum computing}.
\newblock \bibinfo{journal}{\emph{{Quantum}}}  \bibinfo{volume}{2}
  (\bibinfo{date}{Jan.} \bibinfo{year}{2018}), \bibinfo{pages}{49}.
\newblock
\showISSN{2521-327X}
\urldef\tempurl%
\url{https://doi.org/10.22331/q-2018-01-31-49}
\showDOI{\tempurl}


\bibitem[\protect\citeauthoryear{Svore, Cross, Aho, Chuang, and Markov}{Svore
  et~al\mbox{.}}{2004}]%
        {svoretoward}
\bibfield{author}{\bibinfo{person}{K Svore}, \bibinfo{person}{A Cross},
  \bibinfo{person}{A Aho}, \bibinfo{person}{I Chuang}, {and} \bibinfo{person}{I
  Markov}.} \bibinfo{year}{2004}\natexlab{}.
\newblock \showarticletitle{Toward a Software Architecture for Quantum
  Computing Design Tools}.
\newblock \bibinfo{journal}{\emph{Proceedings of the 2nd International Workshop
  on Quantum Programming Languages (QPL)}} (\bibinfo{year}{2004}),
  \bibinfo{pages}{145--162}.
\newblock


\bibitem[\protect\citeauthoryear{Wang, Hu, Sanders, and Kais}{Wang
  et~al\mbox{.}}{2020}]%
        {Wang_2020}
\bibfield{author}{\bibinfo{person}{Y. Wang}, \bibinfo{person}{Z. Hu},
  \bibinfo{person}{B.~C. Sanders}, {and} \bibinfo{person}{S. Kais}.}
  \bibinfo{year}{2020}\natexlab{}.
\newblock \showarticletitle{Qudits and High-Dimensional Quantum Computing}.
\newblock \bibinfo{journal}{\emph{Frontiers in Physics}}  \bibinfo{volume}{8}
  (\bibinfo{date}{Nov} \bibinfo{year}{2020}).
\newblock
\showISSN{2296-424X}
\urldef\tempurl%
\url{https://doi.org/10.3389/fphy.2020.589504}
\showDOI{\tempurl}


\bibitem[\protect\citeauthoryear{Wecker and Svore}{Wecker and Svore}{2014}]%
        {liquid}
\bibfield{author}{\bibinfo{person}{Dave Wecker} {and}
  \bibinfo{person}{Krysta~M. Svore}.} \bibinfo{year}{2014}\natexlab{}.
\newblock \bibinfo{title}{LIQUi$\ket{}$: A Software Design Architecture and
  Domain-Specific Language for Quantum Computing}.
\newblock
\newblock
\urldef\tempurl%
\url{https://doi.org/10.48550/ARXIV.1402.4467}
\showDOI{\tempurl}


\bibitem[\protect\citeauthoryear{Wille, Große, Teuber, Dueck, and
  Drechsler}{Wille et~al\mbox{.}}{2008}]%
        {revlib}
\bibfield{author}{\bibinfo{person}{Robert Wille}, \bibinfo{person}{Daniel
  Große}, \bibinfo{person}{Lisa Teuber}, \bibinfo{person}{Gerhard~W. Dueck},
  {and} \bibinfo{person}{Rolf Drechsler}.} \bibinfo{year}{2008}\natexlab{}.
\newblock \showarticletitle{RevLib: An Online Resource for Reversible Functions
  and Reversible Circuits}. In \bibinfo{booktitle}{\emph{38th International
  Symposium on Multiple Valued Logic (ismvl 2008)}}. \bibinfo{pages}{220--225}.
\newblock
\urldef\tempurl%
\url{https://doi.org/10.1109/ISMVL.2008.43}
\showDOI{\tempurl}


\end{thebibliography}


\end{document}